\documentstyle[11pt,amssymb]{article}
\begin{document}

\newcommand{\be}{\begin{equation}}
\newcommand{\ee}{\end{equation}}
\newcommand{\bes}{\begin{eqnarray}}
\newcommand{\ees}{\end{eqnarray}}
\newcommand{\eens}{\nonumber\end{eqnarray}}


\newcommand{\nl}{\nonumber\\}
\newcommand{\nnl}{\nl[6mm]}
\newcommand{\enl}{\\[6mm]}
\newcommand{\nle}{\nl[-2.5mm]\\[-2.5mm]}
\newcommand{\nlb}[1]{\nl[-2.5mm]\label{#1}\\[-2.5mm]}
\newcommand{\bl}{&&\quad}
\newcommand{\ab}{\allowbreak}

\renewcommand{\/}{\over}
\renewcommand{\d}{\partial}
\newcommand{\eps}{\epsilon}
\newcommand{\dlt}{\delta}
\newcommand{\al}{\alpha}
\newcommand{\la}{\lambda}
\newcommand{\si}{\sigma}
\newcommand{\e}{{\rm e}}
\newcommand{\tr}{{\rm tr}}
\newcommand{\emnr}{\eps^{\mu\nu\rho}}
\newcommand{\emq}{\e^{imq(t)}}

\newcommand{\xmu}{\xi^\mu}
\newcommand{\xnu}{\xi^\nu}
\newcommand{\ynu}{\eta^\nu}
\newcommand{\dmu}{{\d_\mu}}
\newcommand{\dnu}{{\d_\nu}}
\newcommand{\drho}{{\d_\rho}}
\newcommand{\qmu}{{q^\mu}}
\newcommand{\qnu}{{q^\nu}}
\newcommand{\tmu}{{\theta^\mu}}
\newcommand{\tnu}{{\theta^\nu}}
\newcommand{\pmu}{p_\mu}
\newcommand{\pnu}{p_\nu}

\newcommand{\LL}{{\cal L}}
\newcommand{\UU}{{\cal U}}
\newcommand{\J}{{\cal J}}
\newcommand{\G}{{\cal G}}
\newcommand{\HH}{{\cal H}}
\newcommand{\Lxi}{\LL_\xi}
\newcommand{\Leta}{\LL_\eta}

\newcommand{\oj}{{\frak g}}
\newcommand{\ddt}{{d\/dt}}
\newcommand{\qed}{{\hbox{$\  \Box$}}}
\newcommand{\mm}{{\mathbf m}}

\newcommand{\no}[1]{{\,:\kern-0.7mm #1\kern-1.2mm:\,}}

\newcommand{\RR}{{\Bbb R}}
\newcommand{\NN}{{\Bbb N}}

\topmargin 1.0cm

\vspace*{-3cm}
\pagenumbering{arabic}
\begin{flushright}
{\tt math-ph/0002015}
\end{flushright}
\vspace{12mm}
\begin{center}
{\huge Concrete Fock representations of Mickelsson-Faddeev-like 
 algebras}\\[14mm]
\renewcommand{\baselinestretch}{1.2}
\renewcommand{\footnotesep}{10pt}
{\large T. A. Larsson\\
}
\vspace{12mm}
{\sl Vanadisv\"agen 29\\
S-113 23 Stockholm, Sweden}\\
email: tal@hdd.se
\end{center}
\vspace{3mm}
\begin{abstract}
The Mickelsson-Faddeev (MF) algebra can naturally be
embedded in a non-Lie algebra, which suggests that it has no 
Fock representations. The difficulties are due to the inhomogeneous
term in the connection's transformation law.
Omitting this term yields a ``classical MF algebra'',
which has other abelian extensions that do possess Fock modules.
I explicitly construct such modules and the intertwining action of
the higher-dimensional Virasoro algebra.
\end{abstract}

\vskip 1cm

\section{Introduction}
A most challenging and potentially important problem is to construct
the representation theory of local Lie algebras in $N$-dimensional
spacetime. The name indicates that the generators are localized in
spacetime, i.e. that the structure constants are proportional to finitely 
many derivatives of delta functions. This class includes the current 
algebra $map(N,\oj)$, the diffeomorphism algebra $diff(N)$, as well 
as algebras of divergence free,
Hamiltonian, or contact vector fields. When $N=1$, the projective
representations are described by affine Kac-Moody and Virasoro algebras,
respectively, but much less is known in higher dimensions. The reason
for this is the fundamental observation that 
{\em functions and normal ordering are incompatible except in one 
dimension}.
Typically, the classical representations of local Lie algebras are
functions over spacetime, with values in some finite-dimensional 
vector space such as $\oj$ or $gl(N)$ modules. A na\"\i ve strategy
to construct projective representations would be to start from such 
functions, add canonically conjugate momenta, normal order, and hope to 
obtain a realization on Fock space. However, this approach only works in
one dimension; in higher dimensions, new infinities arise.

In view of
this fundamental incompatibility between functions and normal ordering, 
there are two philosophically distinct strategies. One is to keep 
functions and do
something about normal ordering; papers following this route typically
contain the keywords ``further regularization''. The most
ambitious program in this direction has been carried out by Mickelsson
and collaborators, targeting the Mickelsson-Faddeev (MF) algebra
\cite{Fa84,Lang94,Mi85,Mi89,Mi90,Mi91,MR88}. Although representations in 
an abstract sense have been reported, concrete representations (on a 
separable Hilbert space) seem to be missing \cite{Pic89}.

The logical alternative is to keep normal ordering and do something about
functions; more precisely, functions can be replaced by trajectories in
the space of finite jets, which can be viewed as the coefficients of 
truncated Taylor expansions. This route was first entered by
Moody, Eswara-Rao and Yokonoma \cite{EMY92,ERM94,MEY90}, whereas the
geometrical understanding was provided by myself \cite{Lar97,Lar98}.
This approach immediately leads to concrete Fock representations of 
abelian extensions of current algebras. However, these cocycles are not 
of MF type, but rather of the higher-dimensional Kac-Moody type
described by Kassel \cite{Kas85} and rediscovered in \cite{Lar91,MEY90}.
In particular, they involve one-chains rather than three-chains.
Applied to the diffeomorphism algebra, the same method leads to the
higher-dimensional Virasoro algebras of \cite{ERM94,Lar91}.

It is thus natural to ask if the MF algebra could also be represented
using such methods. The answer appears to be negative.
Cederwall et al. \cite{CFNW94} found a natural realization of
a ``classical MF algebra'', where the inhomogeneous term in the 
connection's transformation law has been dropped. However, this term is
not recovered by normal ordering; even worse, it spoils the Jacobi
identities for the realization mentioned above. 
It is shown in the present paper that the classical MF algebra also 
admits other cocycles, but these are of Kac-Moody type.
These new MF-like algebras possess lowest-energy representations,
described in section 3. Moreover, they can be intertwined
with the diffeomorphism algebra, but the extensions are then no longer 
central, since they do not commute with diffeomorphisms.

\section{ Embeddings }
Let $\oj$ be a semisimple finite-dimen\-sional Lie algebra with basis
$J^a$, totally skew structure constants $f^{abc}$, 
Killing metric $\dlt^{ab}$, and brackets $[J^a,J^b] = f^{abc} J^c$. 
As usual, set $d^{abc} = \tr J^a\{J^b,J^c\} \propto \tr J^{(a}J^bJ^{c)}$,
where the trace is evaluated in some representation and paranthesized
indices are symmetrized.
The following identities hold:
$f^{aed}f^{bcd} + f^{acd}f^{bed} + f^{abd}f^{ced} = 0$,
$f^{aed}d^{bcd} + f^{acd}d^{bed} + f^{abd}d^{ced} = 0$, 
$f^{bac} = -f^{abc}$, $f^{bca} = f^{abc}$, 
$d^{bac} = d^{abc}$, and $d^{bca} = d^{abc}$. 

Denote by $\J^a(m)=\exp(im\cdot x)J^a$ the generators of $map(N,\oj)$, 
the algebra of maps  from $\RR^N$ to $\oj$, where
$x = (x^\mu)$ and $m = (m_\mu)$.
Moreover, let $A = A^a_\mu(x) J^a dx^\mu$ be the connection one-form,
with Fourier coefficients $A^a_\mu(m)$.
The Mickelsson-Faddeev (MF) algebra \cite{Fa84,Mi85} 
is the following Lie algebra extension of $map(3,\oj)$:
\bes
[\J^a(m), \J^b(n)] &=& f^{abc} \J^c(m+n) 
 + d^{abc} m_\mu n_\nu \eps^{\mu\nu\rho} A^c_\rho(m+n), \nl
{[}\J^a(m), A^b_\nu(n)] &=& f^{abc} A^c_\nu(m+n) 
 + \dlt^{ab}m_\nu\dlt(m+n), 
\label{MF1}\\
{[}A^\mu(m), A^b\nu(n)] &=& 0,
\eens
where 
\be
\begin{array}{rclll}
\eps^{\mu\nu\rho} &=& +1, &\qquad& 
\hbox{$\mu\nu\rho$ positive permutation of $123$}, \cr
 &=& -1, &\qquad& 
\hbox{$\mu\nu\rho$ negative permutation of $123$}, \cr
&=& 0, &\qquad&
\hbox{otherwise},
\end{array}
\ee
is the totally anti-symmetric symbol in three dimensions.

Set $\HH^{a\mu\nu}(m) = \eps^{\mu\nu\rho} A^a_\rho(m)$. Then the MF
algebra takes the form
\bes
[\J^a(m), \J^b(n)] &=& f^{abc} \J^c(m+n) 
 + d^{abc} m_\mu n_\nu \HH^{c\mu\nu}(m+n), 
\nlb{MF2}
{[}\J^a(m), \HH^{b\mu\nu}(n)] &=& f^{abc} \HH^{c\mu\nu}(m+n) 
 + \dlt^{ab}m_\rho S_3^{\mu\nu\rho}(m+n), 
\eens
and all other brackets vanish. Moreover, $S_3^{\mu\nu\rho}$ is totally 
antisymmetric and subject to the additional condition
\be
m_\rho S_3^{\mu\nu\rho}(m) \equiv 0,
\label{closed}
\ee
which geometrically means that it is a {\em closed} three-chain.
In three dimensions, the unique solution to (\ref{closed}) is 
$S_3^{\mu\nu\rho}(m) = \eps^{\mu\nu\rho}\dlt(m)$, but the present
formulation holds in any number of dimensions $\geq 3$, and in two
dimensions with $S_3^{\mu\nu\rho}(m) \equiv 0$.

Eq. (\ref{MF2}) can be embedded in the following algebra:
\bes
[\J^a(m), \J^b(n)] &=& f^{abc} \J^c(m+n), \nl
{[}\J^a(m), \G^{b\mu}(n)] &=& f^{abc} \G^{c\mu}(m+n), 
\nlb{emb1}
{[}\J^a(m), \HH^{b\mu\nu}(n)] &=& f^{abc} \HH^{c\mu\nu}(m+n) 
 + \dlt^{ab}m_\rho S_3^{\mu\nu\rho}(m+n), \nl
{[}\G^{a\mu}(m), \G^{b\nu}(n)] &=& d^{abc} \HH^{c\mu\nu}(m+n), 
\eens
and all other brackets vanish. Explicitly, this is accomplished by
means of the redefinition
\be
\J^a(m) \mapsto \J^a(m) + m_\mu \G^{a\mu}(m).
\label{redef1}
\ee
In the absence of the closed three-chain $S_3^{\mu\nu\rho}$, this 
embedding was first described by \cite{CFNW94}.

However, there is one big problem with (\ref{emb1}): in the presence of
the three-chain, it is not a Lie algebra, because the following
Jacobi identity fails:
\be
[\J^a(m), [\G^{b\mu}(n), \G^{c\nu}(r)]] + \hbox{cycl.}
= d^{abc} m_\rho S_3^{\mu\nu\rho}(m+n+r) \neq 0.
\ee
I consider this as a strong indication that the MF algebra has no
Fock representations with $S_3^{\mu\nu\rho}(m)$ non-zero. 
At least, it can not be possible to isolate operators 
$\G^{a\mu}(m)$ as in (\ref{redef1}), since that would violate the 
Jacobi identities.

If we skip the three-chain, we obtain the ``classical MF algebra'',
which has another abelian extension:
\bes
[\J^a(m), \J^b(n)] &=& f^{abc} \J^c(m+n) 
 - k \dlt^{ab} m_\rho S_1^\rho(m+n) + \nl
&& + d^{abc} m_\mu n_\nu \HH^{c\mu\nu}(m+n), 
\nlb{MF3}
{[}\J^a(m), \HH^{b\mu\nu}(n)] &=& f^{abc} \HH^{c\mu\nu}(m+n),
\eens
where $S_1^\rho(m)$ is a closed one-form, satisfying
\bes
{[}\J^a(m), S_1^\rho(n)] &=& [\HH^{a\mu\nu}(m), S_1^\rho(n)] = 0, \nle
m_\rho S_1^\rho(m) &\equiv& 0.
\eens
This algebra can be embedded into the following algebra by means of
the same redefinition (\ref{redef1}).
\bes
[\J^a(m), \J^b(n)] &=& f^{abc} \J^c(m+n)
- k \dlt^{ab} m_\rho S_1^\rho(m+n), \nl
{[}\J^a(m), \G^{b\mu}(n)] &=& f^{abc} \G^{c\mu}(m+n), 
\nlb{emb2}
{[}\J^a(m), \HH^{b\mu\nu}(n)] &=& f^{abc} \HH^{c\mu\nu}(m+n), \nl
{[}\G^{a\mu}(m), \G^{b\nu}(n)] &=& d^{abc} \HH^{c\mu\nu}(m+n), 
\eens
where $\G^{a\mu}(m)$ also commutes with the one-chain $S_1^\rho$.

\section{ Fock representations }
Consider the following Lie algebra:
\bes
[J^a(s),J^b(t)] &=& f^{abc}J^c(s)\dlt(s-t) 
 + {k\/2\pi i}\dlt^{ab}\dot\dlt(s-t), \nl
{[}J^a(s),G^{b\nu}(t)] &=& f^{abc}G^{c\nu}(s)\dlt(s-t), \nl
{[}G^{a\mu}(s),G^{b\nu}(t)] &=& d^{abc}H^{c\mu\nu}(s)\dlt(s-t), 
\label{KM} \\
{[}J^a(s),H^{b\mu\nu}(t)] &=& f^{abc}H^{c\mu\nu}(s)\dlt(s-t), \nl
{[}G^{a\mu}(s),H^{b\nu\rho}(t)] &=& 
{[}H^{a\mu\nu}(s),H^{b\si\tau}(t)] = 0,
\eens
where $s,t\in S^1$.
Note that the first relation is the affine Kac-Moody algebra with
central charge $k$.
Moreover, introduce $N$ bosonic oscillators $\qmu(t)$. Then the
following expressions yield a realization of (\ref{emb2}) and thus
of the modified MF algebra (\ref{MF3}).
\bes
\J^a(m) &=& \int dt\ \emq J^a(t), \nl
\G^{a\mu}(m) &=& \int dt\ \emq G^{a\mu}(t), \nle
\HH^{a\mu\nu}(m) &=& \int dt\ \emq H^{a\mu\nu}(t), \nl
S_1^\mu(m) &=& {1\/2\pi}\int dt\ \dot\qmu(t) \emq.
\eens
Moreover, the value of the central charge $k$ is the same in both
formulas.

The problem of finding Fock representations of the modified MF algebra has
thus been reduced to representing (\ref{KM}). This may be done e.g.
by introducing oscillators $\phi^a(t)$, $\psi^{a\mu}(t)$, and
$\zeta^{a\mu\nu}(t)$, together with their canonical conjugate momenta
$\bar\phi^a(t)$, $\bar\psi^a_\mu(t)$, and $\bar\zeta^a_{\mu\nu}(t)$.
Moreover, $\zeta^{a\mu\nu}(t)$ and $\bar\zeta^a_{\mu\nu}(t)$ are 
assumed to be symmetric in $\mu\nu$.
The canonical commutation relation read
\bes
[\bar\phi^a(s), \phi^b(t)] &=& \dlt^{ab}\dlt(s-t), \nl
{[}\bar\psi^a_\mu(s), \psi^{b\mu}(t)] &=& 
\dlt^{ab}\dlt^\nu_\mu\dlt(s-t), 
\label{Fock}\\
{[}\bar\zeta^a_{\mu\nu}(s), \zeta^{b\si\tau}(t)] &=&
\dlt^{ab}\dlt^{(\si}_\mu\dlt^{\tau)}_\nu\dlt(s-t),
\eens
and all other brackets vanish. Then the following operators
\bes
J^a(t) &=& f^{abc} ( \no{\phi^c(t)\bar\phi^b(t)}
 +  \no{\psi^{c\mu}(t)\bar\psi^b_\mu(t)}
 + \no{\zeta^{c\mu\nu}(t)\bar\zeta^b_{\mu\nu}(t)} ), \nl
G^{a\mu}(t) &=& f^{abc} \psi^{c\mu}(t)\bar\phi^b(t)
 + d^{abc} \zeta^{c\mu\nu}(t)\bar\psi^b_\nu(t), 
\label{JGH}\\
H^{a\mu\nu}(t) &=& f^{abc} \zeta^{c\mu\nu}(t)\bar\phi^b(t),
\eens
satisfy (\ref{KM}). The double dots in the first expression indicate
standard one-dimensional normal ordering with respect to frequency.

\section{ Diffeomorphisms }
The algebra (\ref{emb2}), and thus also the modified MF algebra 
(\ref{MF3}),
admits an intertwining action of an extension of the diffeomorphism
algebra $diff(N)$. The additional brackets read
\bes
[\LL_\mu(m), \LL_\nu(n)] &=& n_\mu \LL_\nu(m+n) - m_\nu \LL_\mu(m+n) \nl
&&+(c_1 m_\nu n_\mu + c_2 m_\mu n_\nu) m_\rho S_1^\rho(m+n), \nl
{[}\LL_\mu(m), \J^a(n)] &=& n_\mu \J^a(m+n), \nl
{[}\LL_\mu(m), \G^{a\nu}(n)] &=& n_\mu \G^{a\nu}(m+n)
 + \dlt^\nu_\mu m_\rho\G^{a\rho}(m+n), 
\label{diff} \\
{[}\LL_\mu(m), \HH^{a\nu\rho}(n)] &=& n_\mu \HH^{a\nu\rho}(m+n) \nl
 &&+ \dlt^\nu_\mu m_\si \HH^{a\si\rho}(m+n) 
 + \dlt^\rho_\mu m_\si \HH^{a\nu\si}(m+n), \nl
{[}\LL_\mu(m), S_1^\nu(n)] &=& n_\mu S_1^\nu(m+n)
 + \dlt^\nu_\mu m_\rho S_1^\rho(m+n),
\eens
where $\LL_\mu(m)$ are the $diff(N)$ generators and the cocycles
multiplied by $c_1$ and $c_2$ were first found by Eswara-Rao and Moody
\cite{ERM94} and myself \cite{Lar91}, respectively. For the classification
of $diff(N)$ cocycles, see \cite{Dzhu96} and \cite{Lar00}. To construct
a representation of (\ref{diff}), introduce $N$ oscillators $p_\mu(t)$
which are the canonical momenta of $\qmu(t)$, i.e.
\be
[p_\mu(s), \qnu(t)] = \dlt^\nu_\mu \dlt(s-t), \qquad
[p_\mu(s), p_\nu(t)] = [\qmu(s), \qnu(t)] = 0.
\ee
The $diff(N)$ generators have the realization
\be
\LL_\mu(m) = \int dt\ \Big( -i\no{ \emq p_\mu(t)} 
 + m_\nu \emq T^\nu_\mu(t) \Big), 
\ee
where $T^\mu_\nu(t)$ are the generators of the Kac-Moody algebra
$\widehat{gl(N)}$. The relevant relations read
\bes
[T^\mu_\nu(s), T^\si_\tau(t)] &=&
(\dlt^\si_\nu T^\mu_\tau(s) - \dlt^\mu_\tau T^\si_\nu(s) )\dlt(s-t) \nl
&& - {1\/2\pi i} (k_1 \dlt^\mu_\tau\dlt^\si_\nu + 
 k_2 \dlt^\mu_\nu\dlt^\si_\tau) \dot\dlt(s-t), \nl
{[}T^\mu_\nu(s), q^\rho(t)] &=& 
{[}T^\mu_\nu(s), p_\rho(t)] = 0, \\
{[}T^\mu_\nu(s), J^a(t)] &=& 0, \nl
{[}T^\mu_\nu(s), G^{a\si}(t)] &=& \dlt^\si_\nu G^{a\mu}(s)\dlt(s-t), \nl
{[}T^\mu_\nu(s), H^{a\si\tau}(t)] &=&
(\dlt^\si_\nu H^{a\mu\tau}(s) + \dlt^\tau_\nu H^{a\si\mu}(s)) \dlt(s-t),
\eens
and the abelian charges in (\ref{diff}) take the values
$c_1=1+k_1$, $c_2 = k_2$.
$\widehat{gl(N)}$ acts on the same Fock space as the operators in
(\ref{JGH}), by means of the following expression:
\bes
T^\mu_\nu(t) &=& \dlt^\mu_\nu( \no{\phi^a(t)\bar\phi^a(t)} +
\no{\psi^{a\si}(t)\bar\psi^a_\si(t)} 
+\no{\zeta^{a\si\tau}(t)\bar\zeta^a_{\si\tau}(t)}) + \nle
&&+\no{\psi^{a\mu}(t)\bar\psi^a_\nu(t)} 
+\no{\zeta^{a\mu\rho}(t)\bar\zeta^a_{\nu\rho}(t)}.
\eens

As was noted in \cite{Lar98}, we can actually represent a larger
algebra on the same Fock space. Namely, there is a natural action of
an additional $diff(1)$ algebra, which classically commutes with
both $diff(N)$ and the MF algebra. Geometrically, this algebra describes
reparametrizations of the observer's trajectory.

\section{ Conclusion }
Originally, the present study had two goals: to construct projective Fock 
representations of the classical MF algebra, and to find representations
where the cocycle had precisely the MF form (\ref{MF2}). 

The first goal was easily reached using the formalism developped
in \cite{Lar97,Lar98,Lar99}. It is clear 
that much more complicated representations can be written down along 
the same lines. Geometrically, the oscillators can be viewed as
zero-jets, i.e. the value of fields like $\phi^a(x)$ on the 
observer's trajectory $x^\mu = \qmu(t)$. One can generalize to $p$-jets, 
with basis $\d_{\nu_1}\ldots\d_{\nu_r}\phi^a(q(t)))$ for all $r\leq p$. 
This is a genuine $N$-dimensional object which probes not only the value 
of the fields along the trajectory, but also finitely many transverse 
derivatives. On the other hand, the presence of the MF term
$\HH^{a\mu\nu}(m)$ is quite uninteresting, since it can
be disentangled using (\ref{redef1}). The interesting quantum (normal
ordering) effect
is the Kac-Moody extension for the $\J\J$ bracket in (\ref{emb2}).

However, my second goal failed. Indeed, the fact that the true MF 
algebra (\ref{MF2}) can be naturally embedded into the non-Lie algebra 
(\ref{emb1}) is a serious obstruction against the existence of Fock
modules. I am convinced that the technique of combining normal ordering 
with jet space trajectories can only produce one-chain (Kac-Moody)
cocycles. This is a serious problem because this technique has so far 
been the only viable method to produce concrete Fock modules in more 
than one dimension. Two conclusions are possible: either the MF algebra
lacks Fock modules altogether, or it points to a new type of 
representation theory which is not yet understood.


\begin{thebibliography}{99}

\bibitem{Fa84} L. D. Faddeev, Phys. Lett. {\bf 145B} (1984) 81.

\bibitem{CFNW94} M. Cederwall, G. Ferretti, B.E.W. Nilsson and
A. Westerberg, Nucl. Phys. {\bf B424} (1994) 97.

\bibitem{Dzhu96} A. Dzhumadil'daev,
  Z. Phys. C {\bf 72} (1996) 509.

\bibitem{EMY92} S. Eswara Rao, R.V. Moody and T. Yokonuma,
  Nova J. of Algebra and Geometry {\bf 1} (1992) 15.

\bibitem{ERM94} S. Eswara Rao and R.V. Moody,
  Comm. Math. Phys. {\bf 159} (1994) 239.

\bibitem{Kas85} C. Kassel,
  J. Pure and Appl. Algebra {\bf 34} (1985) 256.

\bibitem{Lang94} E. Langmann, 
  Comm. Math. Phys. {\bf 162} (1994) 1.

\bibitem{Lar91} T.A. Larsson,
  J. Phys. A. {\bf 25} (1992) 1177.

\bibitem{Lar97} T.A. Larsson, 
  Comm. Math. Phys. {\bf 201} (1999) 461.

\bibitem{Lar98} T.A. Larsson, 
  {\tt math-ph/9810003} (1998).

\bibitem{Lar99} T.A. Larsson, 
  {\tt math-ph/9908028} (1999).

\bibitem{Lar00} T.A. Larsson, 
  {\it Extensions of diffeomorphism and current algebras},
  {\tt math-ph/0002016} (2000).

\bibitem{MEY90} R.V. Moody, S. Eswara Rao and T. Yokonoma,
  Geom. Ded. {\bf 35} (1990) 283.

\bibitem{Mi85} J. Mickelsson,
  Comm. Math. Phys. {\bf 97} (1985) 361.

\bibitem{Mi89} J. Mickelsson,
  {\it Current algebras and groups},
  Plenum Monographs in Nonlinear Physics, London: Plenum Press, 1989.

\bibitem{Mi90} J. Mickelsson,
  Comm. Math. Phys. {\bf 127} (1990) 285.

\bibitem{Mi91} J. Mickelsson,
  Lett. Math. Phys. {\bf 28} (1993) 97.

\bibitem{MR88} J. Mickelsson and S. Rajeev, 
  Comm. Math. Phys. {\bf 116} (1988) 365.

\bibitem{Pic89} Pickrell,
  Comm. Math. Phys. {\bf 123} (1989) 617.

\bibitem{Wes96} A. Westerberg,
  {\tt hep-th/9612167}, (1996).

\end{thebibliography}
\end{document}